\documentclass[11pt,twoside]{article}

\usepackage{asp2006}
\usepackage{epsf}
\usepackage{lscape}
\usepackage{graphicx}

\begin{document}
\title{Resolving the BLR in NGC~3783}   %%% Fill in title
\author{P.~Lira$^{1}$, M.~Kishimoto$^{2}$, A.~Robinson$^{3}$, S.~Young$^{4}$, D.~Axon$^{3}$, M.~Elvis$^{5}$, A.~Lawrence$^{6}$ \& B.~Peterson$^{7}$}   %%% Fill in author names
\affil{
$^{1}$Universidad de Chile, Casilla 36D, Santiago, Chile\\
$^{2}$MPIfR, Auf dem H\"ugel 69, 53121 Bonn, Germany\\
$^{3}$RIT, 54 Lomb Memorial Dr., Rochester, NY 14623\\
$^{4}$University of Hertfordshire, Hatfield AL10 9AB, UK\\
$^{5}$Center for Astrophysics, 60 Garden Street, Cambridge, MA02138, USA\\
$^{6}$University of Edinburgh, Blackford Hill, Edinburgh EH9 3HJ, UK\\
$^{7}$OSU, 140 W. 18th Avenue, Columbus, OH 43210-1173, USA
}    %%% Fill in author affiliations

\begin{abstract} %%% Abstract to run on from here.
We present results from very high signal-to-noise spectropolarimetric
observations of the Seyfert 1 galaxy NGC~3783. Position Angle (PA)
changes across the Balmer lines show that the scatterer is {\it
resolving\/} the Broad-Emission Line Region (BLR). A broad component
seen in polarized light and located bluewards from the H$\beta$ line
very likely corresponds to HeII$\lambda4686$. The lack of PA changes
across this line suggests that the region responsible for this
emission appears to the scatterer as unresolved as the continuum
source, in agreement with the stratified BLR structure determined from
reverberation mapping.

\end{abstract}

%%% MAIN BODY OF TEXT GOES HERE. CONSULT "INSTRUCTIONS FOR AUTHORS USING
%%% LATEX2E MARKUP", SECTIONS 2.3-2.6 FOR HELP WITH EQUATIONS, FIGURES,
%%% AND TABLES.

%\section{}   %%% Top level section head (remove "%" symbol)
%\subsection{}   %%% Second level section head (remove "%" symbol)
%\subsubsection{}   %%% Lowest level section head (remove "%" symbol)
%\section*{}    %%% Unnumbered top level section head (remove "%" symbol)
%\subsection*{}   %%% Unnumbered second level section head (remove "%" symbol)

\section{Observations}

We have obtained very high S/N spectropolarimetric observations of
NGC~3783 using the VLT and the 3.6m telescope at La Silla in 2006,
with total exposure times of 3.4 and 6.2 hours, respectively. The data
were reduced following Miller, Robinson \& Goodrich (1988) and special
care was taken to correct for the interstellar polarization in our
Galaxy along the line of sight towards NGC~3783.

\section{Main Results}

The left panel in Figure 1 shows the total flux and PA of the
polarized emission in the 4000-5000 \AA\ range. Strong PA changes are
coincident with the broad Balmer emission lines. This is consistent
with near-field scattering, i.e., with the scatterer being close
enough to the Balmer emitting region to {\it resolve\/} it. These PA
changes, however, are not consistent with a simple, pure rotational
motion of the BLR as seen by an equatorial scattering medium, which
predicts a horizontal {\it S-shaped\/} PA swing, as already observed
in several other Seyfert 1 galaxies, and modeled by Smith et
al.~(2005). Instead, a {\it M-shaped\/} pattern is seen in all Balmer
lines (also clear in H$\alpha$, which is not shown here). The Balmer
lines also show a narrow dip in polarized flux which is blue-shifted
from the position of the emission peak seen in total flux (central
panel in Figure 1). In a forthcoming paper we will present detailed
modeling of our data.

The HeII$\lambda4686$ line, very conspicuous in total flux, does not
show the features observed in the polarized flux of the Balmer lines,
and essentially no PA change. The right panel in Figure 1 compares the
H$\alpha$ and H$\beta$ profiles in velocity space. An excess is
clearly seen extending from the blue wing of H$\beta$, which is
coincident with the position of the HeII$\lambda4686$ emission
line. We therefore interprete this excess as polarized emission from
the broad HeII$\lambda4686$ line. The lack of a PA change across the
HeII$\lambda4686$ line strongly suggests a smaller solid angle
subtended by the emitting region as seen by the scatterer when
compared with the Balmer emitting region.

\section{Discussion}
 
We have found clear evidence that the scatterer is resolving the
Balmer emitting region in NGC~3783. The geometry and kinematics will
be explored in a forthcoming paper which will model the
spectropolarimetric observations.

Reverberation mapping results show clear evidence for a stratified BLR
which is also consistent with Virial motion of the BLR (Onken
\& Peterson, 2002). We have found strong evidence that the high
ionization HeII$\lambda4686$ line is produced in a region much more
compact than that producing the Balmer lines, in good agreement with
the idea of a stratified BLR.

\setlength{\unitlength}{1cm}
\begin{figure}
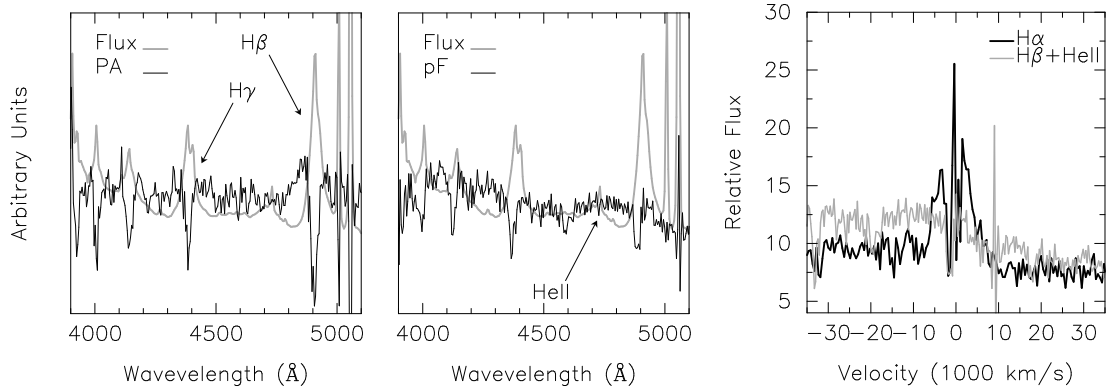

\begin{picture}(15,5){
\put(-0.5,0){\includegraphics[scale=0.5]{p_lira_fig1.ps}}
\put(9,0){\includegraphics[scale=0.5]{p_lira_fig2.ps}}}
\end{picture}
\caption{{\it Left and central panels:} Comparison between total flux
and polarization position angle (PA), and total and polarized flux
spectra. {\it Right panel:} H$\alpha$ and H$\beta$ emission line
profiles in velocity space. We identify the excess seen in the blue
wing of H$\beta$ as emission from the HeII$\lambda4686$ line.}
\end{figure}

\acknowledgements 
PL acknowledges support by Fondap project \#15010003 and Fondecyt
project \#1040719.

%%% THE BIBLIOGRAPHY
%%%
%%% CONSULT SECTION 3 OF "INSTRUCTIONS FOR AUTHORS" FOR HOW TO USE NATBIB.
%%% AUTHORS ARE ENCOURAGED TO USE EITHER THE "THEBIBLIOGRAPY" ENVIRONMENT
%%% BY UNCOMMENTING (DELETING THE "%" SYMBOL) THE COMMANDS BELOW, OR BY
%%% USING THE BIBTEX ENVIRONMENT. TO FIND OUT WHICH IS APPLICABLE TO YOUR
%%% CONTRIBUTION, CONSULT THE VOLUME EDITORS FOR YOUR PROCEEDINGS.
%%%

\end{document}